\documentclass[]{spie}  

\newcommand{\pseudodot}{{\lower 2.4pt\hbox{$\cdot$}}}

\usepackage{amsmath}
\usepackage{graphicx}
\usepackage[caption=false]{subfig}
\usepackage{float}
\usepackage{multirow,array,booktabs}

\DeclareUnicodeCharacter{2212}{-}
\newcommand{\caltech}{Department of Astronomy, California Institute of Technology, Pasadena, CA 91125, USA}
\newcommand{\jpl}{Jet Propulsion Laboratory, California Institute of Technology, 4800 Oak Grove Dr., Pasadena, CA 91109, USA}
\usepackage{amsmath,amsfonts,amssymb}
\usepackage{graphicx}
\usepackage{threeparttable}

\providecommand{\adsurl}[1]{}

\usepackage[colorlinks=true, allcolors=blue]{hyperref}
\usepackage{amsmath}
\title{Laboratory demonstration of low order wavefront control using light reflected off the vortex coronagraph}

\author[a]{Clarissa R. Do Ó}
\author[a]{Kane Sjoberg}
\author[a]{Luke Lamitina}
\author[b]{Susan Redmond}
\author[a,b]{Dimitri Mawet}
\author[b]{Jorge Llop-Sayson}
\author[c]{Siria Alicata}

\affil[a]{\caltech}
\affil[b]{\jpl}
\affil[c]{Trinity College Dublin, the University of Dublin, Dublin 2, D02 PN40, Ireland}

\authorinfo{Further author information: (Send correspondence to C.D.O) \\ E-mail:cdoo@caltech.edu}

\authorinfo{Further author information: (Send correspondence to Clarissa R. Do Ó)\\Clarissa Do Ó: E-mail: cdoo@ucsd.edu}

\begin{document} 
\maketitle

\begin{abstract}
The Astro2020 Decadal Survey identified exoplanet imaging as a high priority for the Habitable Worlds Observatory (HWO), which will require imaging and characterizing exo-Earths at contrasts of $1\times10^{-10}$. The vector vortex coronagraph (VVC) is a leading candidate architecture for this task due to its small inner working angle and high throughput. Using light reflected from the VVC for wavefront sensing and control provides a potential path to improving robustness to residual wavefront errors and minimizing contrast degradation. The High Contrast and Spectroscopy Testbed (HCST) at Caltech's Exoplanet Technology Laboratory (ETLab) is an in-air coronagraphic testbed that has demonstrated $1\times10^{-8}$ broadband contrast between 3.5--10~$\lambda/D$ using a deformable mirror (DM) and a charge 8 VVC. As a first step toward full reflected light wavefront control, we present the first laboratory demonstration of a low order wavefront sensing and control (LOWFS) loop using light reflected from the VVC. This work is enabled by HCST's recent software architecture upgrade to CATKit2, a Python-based and service-oriented framework that allows complex lab routines to run fast and concurrently. We use CATKit2 to implement phase retrieval, electric field conjugation (EFC), and the LOWFS control loop routines on HCST. Using a Gerchberg--Saxton phase retrieval routine, we reduce the science camera wavefront error from 71.6 to 7.9 nm RMS, a $>$9$\times$ improvement. Over a 12 hour run with LOWFS tip-tilt tracking, we find that PSF drift is strongly correlated with bench temperature ($r>0.88$). Closing the LOWFS tip-tilt loop suppresses low frequency ($<$1 Hz) drift by more than two orders of magnitude, maintaining pointing precision at $<0.005~\lambda/D$. When EFC and LOWFS are run concurrently, the closed LOWFS loop maintains a dark hole contrast of $\sim$5$\times10^{-8}$ over one hour, while the open loop case drifts by $\sim$0.2--0.4~$\lambda/D$ and degrades in contrast by more than one order of magnitude. These results demonstrate that tip-tilt aberrations can be sensed and stabilized using light reflected from a VVC, establishing a foundation for future reflected light wavefront sensing and control architectures for high contrast coronagraphy.

\end{abstract}
 
\section{INTRODUCTION}
The \href{https://www.nationalacademies.org/projects/DEPS-BPA-18-01}{Astro2020} decadal survey has recommended that NASA begin the development of the Habitable Worlds Observatory (HWO), a mission in the ultraviolet, infrared, and optical wavelengths that will search for Earth-like worlds and signs of life around other stars. In light of this recommendation, HWO will have the capability to image Earth-like planets around Sun-like stars. As a technical requirement, HWO will need to achieve a contrast level of $1\times10^{-10}$ at 3$\lambda/D$. Its scientific promise to directly detect and characterize Earth-like planets in reflected light depends critically on selecting a coronagraph architecture capable of both deep starlight suppression and long-term stability. \par
Among the leading candidates is the vector vortex coronagraph (VVC), a spiral phase mask that redistributes starlight outside the pupil while allowing nearby planetary light to pass through \cite{Mawet2005, Foo2005}. The vortex architecture is particularly attractive because it enables small inner working angles (IWA), permitting observations closer to the star than many alternative designs \cite{LlopSayson2024}, as well as a high off-axis throughput. However, coronagraphs are highly sensitive to wavefront error, especially tip-tilt errors that allow residual starlight to leak through and overwhelm the planet signal. To reach HWO’s required capability of imaging planets $10^{10}\times$ fainter than their stars, the telescope must continuously measure and correct for wavefront aberrations\cite{Mawet2010}. 
\par
Although in theory a wavefront control architecture can help stabilize the performance of VVCs, this approach needs to be experimentally validated. In this work, we use the High Contrast and Spectroscopy Testbed (HCST) at Caltech's Exoplanet Technology Laboratory (ETLab) to demonstrate for the first time the implementation of a tip-tilt (pointing) control loop using the light reflected off the VVC for wavefront sensing and control. 

\section{Motivation}
The contrast performance of coronagraphs is highly sensitive to wavefront errors, in particular tip-tilt (pointing) upstream of the focal plane mask (FPM) in the optical path \cite{Kenworthy2025, Guyon2009}. There are various causes for such errors, including thermal drift, vibrations, mechanical instabilities, and, in in-air testbeds, temperature and humidity variations. These errors lead to a decreased contrast performance and stability, emerging as speckles in the final coronagraphic image. For that reason, there have been many strategies to mitigate these speckles and improve coronagraphic performance such that fainter and closer-in planets can be detected. To mitigate these effects and improve sensitivity to faint, close-in exoplanets, a variety of low-order wavefront sensing (LOWFS) strategies have been developed. These approaches differ primarily in where the sensing is performed within the optical path.
\par

Wavefront sensing implemented upstream of the coronagraph is more susceptible to non-common path aberrations between the sensing and science channels \cite{Kenworthy2025}. Alternatively, sensing can be performed using light rejected or modified by the coronagraph itself, such as through reflection of the FPM \cite{Guyon2009} or by utilizing diffracted starlight at the Lyot stop \cite{Singh2014}. These latter approaches reduce non-common path errors by probing the wavefront closer to the coronagraph, but rely on signals that have been transformed by the FPM and therefore require calibration to interpret.
\par
Among these approaches, LOWFS architectures based on reflection at the FPM have demonstrated strong performance due to their relatively simple implementation and reduced non-common path errors \cite{Guyon2009}. This strategy has therefore been adopted for the Coronagraph Instrument (CGI) aboard the Nancy Grace Roman Space Telescope \cite{Riggs2025}. However, such architectures have primarily been developed for Lyot coronagraphs and have not yet been extensively demonstrated with VVCs. Despite their theoretically perfect on-axis starlight rejection, VVCs are still sensitive to wavefront aberrations, particularly at low topological charge where the inner working angle is smallest \cite{Kenworthy2025}. This makes them especially promising for detecting close-in exoplanets, such as exo-Earths, while simultaneously motivating the need for robust LOWFS architectures tailored to their optical behavior.

Caltech's HCST is ideally equipped to experimentally validate a reflection LOWFS+VVC architecture, as a LOWFS setup at the reflection off the VVC was introduced on the testbed in 2024 \cite{Bertrou-Cantou2024} after its optical redesign in 2023 \cite{Bertrou-Cantou2023}.

\section{Experimental Setup}
HCST is an in-air testbed at Caltech's ETLab. Its optical re-design is presented in \cite{Bertrou-Cantou2023}. Here we present HCST's components relevant to this work and our implementation of a control loop and an electric field conjugation (EFC) algorithm with the new software architecture, CATKit2. 

\subsection{HCST's Optical Design}
\begin{figure} [ht!]
   \begin{center}
   \begin{tabular}{c}
   \includegraphics[width = \textwidth]{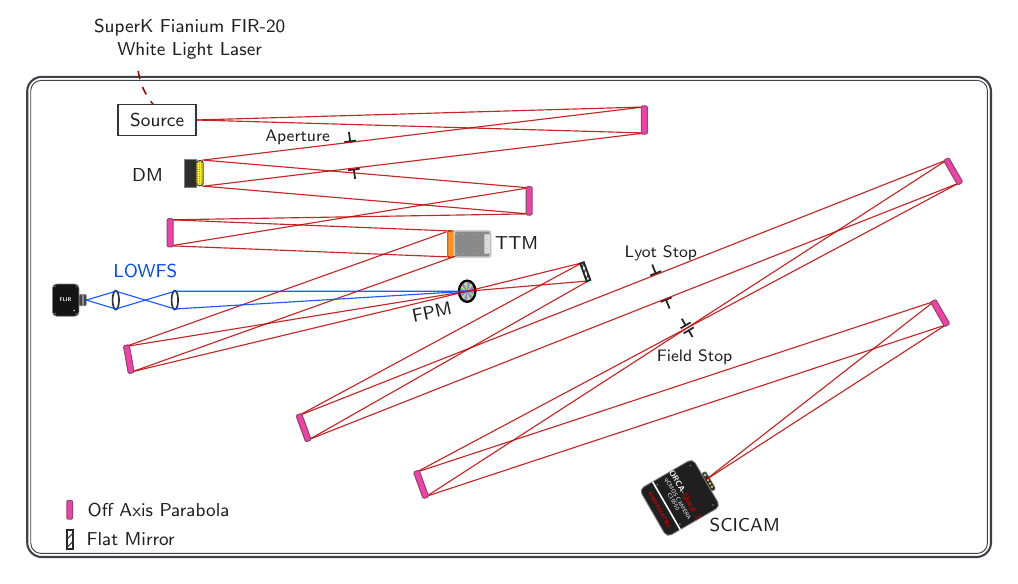}
   \end{tabular}
   \end{center}
   \caption[example] 
   {\label{fig:hcst} HCST's optical layout at Caltech's ETLab. Pink rectangles represent off axis parabolic (OAP) mirrors. The 34x34 DM is 8 cm away from where the first pupil is defined. The pupil is defined by an aperture with a physical size of 10 mm. The TTM is coupled to the LOWFS readings for controlling the beam that reaches the FPM, currently a charge 8 VVC. The beam reaches the Lyot stop, which transmits 84\% of the entrance beam, a Field Stop to isolate the region relevant to dark hole digging, and a low noise qCMOS ORCA Quest 2 science camera (SCICAM) manufactured by Hamamatsu.}
   \end{figure}

The schematic of HCST is presented in Figure \ref{fig:hcst}. The source assembly is a white light laser from NKT Photonics connected to a tunable filter. Other important components of HCST include a Boston Micromachines 34x34 deformable mirror (DM), a tip-tilt mirror (TTM) connected to a two axis platform by PI, a charge 8 liquid crystal polymer VVC, a Lyot stop that transmits 84\% of the entrance beam, and a Hamamatsu Orca-Quest 2 qCMOS science camera (SCICAM). The LOWFS arm consists of two Thorlabs lenses with an intermediate focal plane and a Teledyne FLIR camera positioned in the final focal plane. The operating wavelength of HCST is 780 nm. 

The LOWFS layout\cite{Bertrou-Cantou2024} is designed to have a magnification factor that guarantees the point spread function (PSF) is sampled above the Nyquist sampling and an intermediate focal plane such that a Zernike phase mask can be placed there in the future, with room to slide the FLIR camera to a pupil plane for full Zernike wavefront sensing. This layout will be tested and presented in future work. As of now, HCST can perform only tip-tilt sensing and control of the light reflected off the VVC. These lower order aberrations are usually the largest contributors to the degrading of coronagraphic performance.

\subsection{CATKit2 Implementation}
The previously used software architecture on HCST was the FAst Linearized COronagraph optizimer (FALCO) \cite{Riggs2018}, which is MATLAB based. The main motivation for replacing the software architecture was the lack of support for running parallel control loops and its overall speed. The new software architecture, which is written in Python, is CATKit2\cite{por_2026}. CATKit2 uses a service oriented architecture, where each service is designed to operate a specific piece of hardware on the testbed (e.g. the TTM, SCICAM). Since Summer 2025, every piece of hardware has been integrated into CATKit2, including not only the detectors and mirrors, but also linear stages (Zaber and Thorlabs) used to align the optical components. The services operate as wrappers for each component's SDK and make a user friendly frontend for operating the testbed. This facilitates the development of routine operations such as phase retrieval, the LOWFS PID control loop, and dark hole digging with EFC, outlined in detail in the following sections.

\subsubsection{Phase Retrieval}

Coronagraphy relies on minimizing wavefront aberrations to produce a clean
PSF for effective starlight suppression. To minimize residual aberrations
from optical misalignments and imperfections on HCST, we run a phase retrieval
routine between the DM and SCICAM to generate a DM ``flat map'' that is always applied
prior to routine operations. We implement this in CATKit2 using a
Gerchberg--Saxton (GS) algorithm \cite{Gerchberg1972}.

GS is an iterative method for reconstructing wavefront phase from intensity
measurements at the focal and pupil planes. HCST has a filter wheel in front of the SCICAM that allows for pupil imaging. GS carries a
sign ambiguity, since $-\phi$ and $+\phi$ produce identical intensities. We
break the ambiguity by repeating the retrieval at four defocus positions by moving the SCICAM on a linear stage and taking the
median of the recovered phase. Starting from the measured pupil
amplitude $A_p$ and a zero initial phase, each iteration propagates the field
(with the current phase guess plus the known defocus $a_4 Z_4$) to the focal
plane,
\begin{equation}
    E_f = \mathcal{F}\left[A_p \, e^{i(\phi^{(k)} + a_4 Z_4)}\right],
\end{equation}
replaces the modeled focal plane amplitude with the measured PSF amplitude
while keeping the modeled phase,
\begin{equation}
    E_f' = \sqrt{I_{\mathrm{meas}}} \, e^{i\,\phi_f}, \quad \phi_f = \arg(E_f),
\end{equation}
and propagates back to the pupil, removing the defocus term,
\begin{align}
    E_p &= \mathcal{F}^{-1}\left[E_f'\right] \, e^{-ia_4 Z_4}, \\
    \phi^{(k+1)} &= \arg(E_p).
\end{align}
The pupil amplitude is then reimposed and the process repeats. We run 300 GS iterations per defocus image and
apply the resulting correction over 6 rounds at a gain of $0.7$ to avoid
overcorrection.

\begin{figure*}[ht!]
  \begin{center}
    \includegraphics[width=\textwidth]{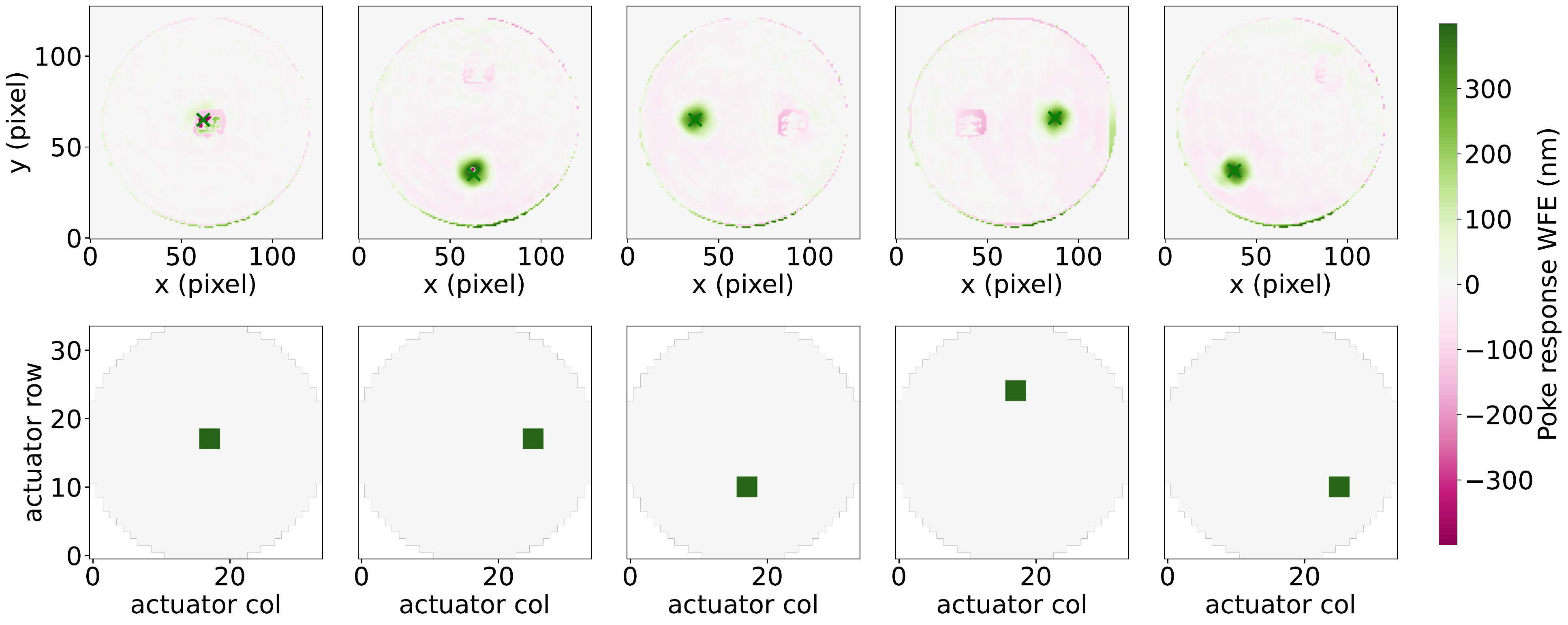}
    \caption{Mapping the pupil (top) to DM (bottom) coordinates.}
    \label{fig:mappin}
  \end{center}
\end{figure*}

GS recovers a phase map on a pupil grid, but the DM correction must be applied
on the $34\times34$ actuator grid, whose orientation does not match the SCICAM.
To map between the two, we poke a $3\times3$ block of actuators at five positions
across the DM and locate the pokes
with GS (Figure~\ref{fig:mappin}). The five DM--pupil coordinate pairs are fit
with a 2D affine transform,
\begin{align}
    x_{\mathrm{pup}} &= 0.108\,c + 3.629\,r - 1.2, \\
    y_{\mathrm{pup}} &= -3.558\,c + 0.057\,r + 125.2,
\end{align}
where $(r, c)$ are actuator row and column indices and
$(x_{\mathrm{pup}}, y_{\mathrm{pup}})$ are pixel coordinates on the
$128\times128$ GS phase grid. The mapping gives a scale of 3.59 pupil pixels
per actuator and a rotation of $-88^\circ$ between the DM and pupil coordinate
systems, set by the orientation of the DM relative to the SCICAM.

With this calibration, the GS phase is sampled at each actuator's mapped
position by linear interpolation,
\begin{equation}
    \phi_{\mathrm{DM}}(r, c) = \phi_{\mathrm{GS}}\!\left(y_{\mathrm{pup}}(r,c),\; x_{\mathrm{pup}}(r,c)\right),
\end{equation}
and converted to a surface height. We deconvolve the DM actuator’s influence function from the target surface to recover the actuator commands that reproduce the desired shape. 
\subsubsection{PID Tip-Tilt Control Loop} \label{pid}
\begin{figure} [ht!]
   \begin{center}
   \begin{tabular}{c}
   \includegraphics[width = 0.9\textwidth]{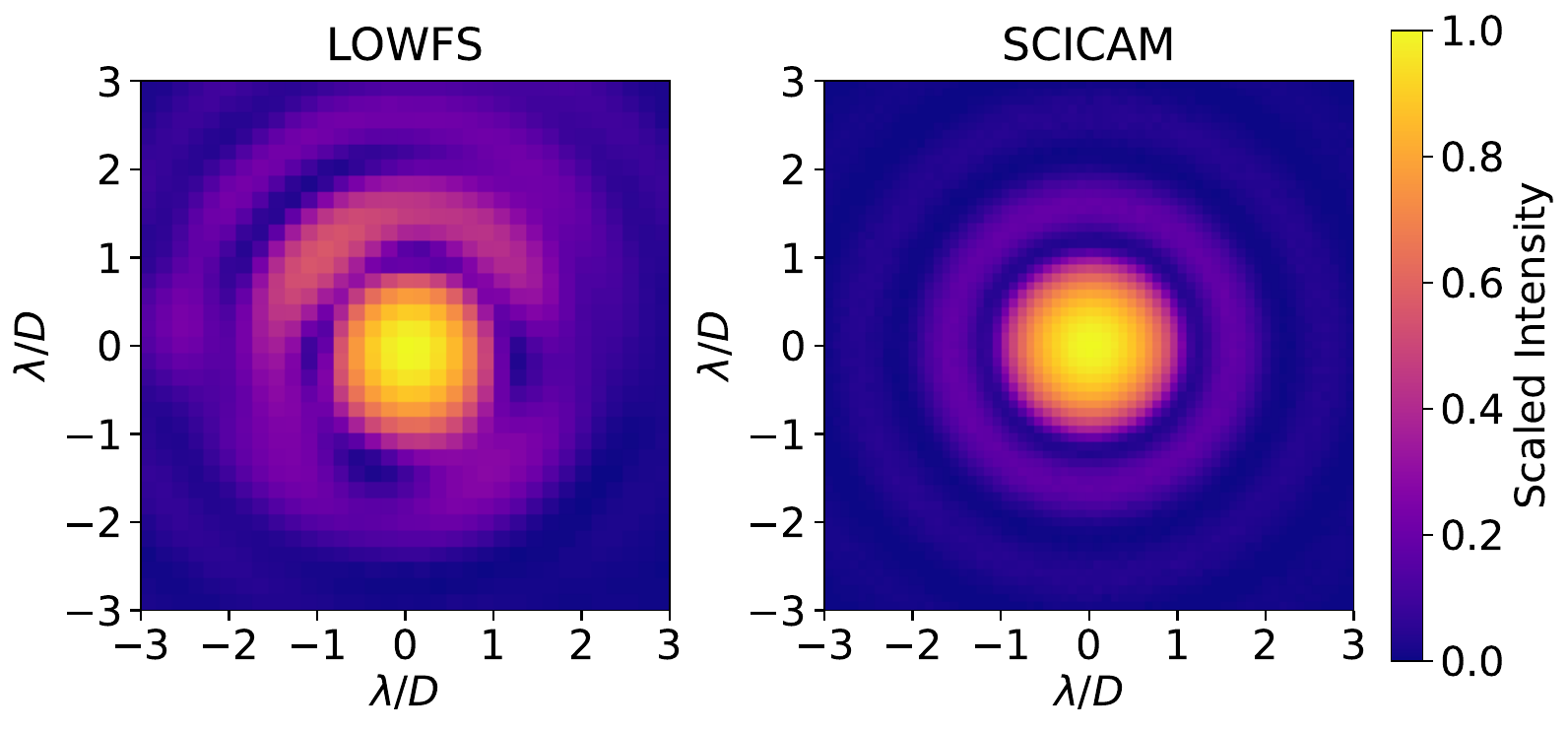}
   \end{tabular}
   \end{center}
   \caption[example] 
   {\label{fig:psfs} Left: The PSF image for the LOWFS system, which is obtained via the reflection of the PSF off the VVC (blue light path in Figure \ref{fig:hcst}). It is sampled at 5.2 pixels/$\lambda$/D. The asymmetries arise due to aberrations caused by our two lens system. Right: PSF image from our SCICAM. It is oversampled at about 10.3 pixels/$\lambda$/D.}
   \end{figure}

The LOWFS re-images the PSF incident on the VVC onto a FLIR camera (see Figure \ref{fig:psfs}), allowing us to monitor residual tip-tilt jitter and drift of the PSF relative to the vortex center (see Figure \ref{fig:hcst}, blue optical path). The LOWFS sampling is approximately 5.2 pixels per $\lambda/D$. The TTM, located upstream of the VVC, is used to correct this motion by steering the PSF back to the VVC center.
   
The PSF position is measured on each frame using both Gaussian and center of mass centroiding. A reference centroid is first established when the PSF is aligned to the VVC center. We assess the behavior of the centroid in both open loop (tracking) and closed loop (controlling).  During the closed loop operation, the measured centroid is compared against this reference position to form an error signal,
\begin{equation}
    \mathbf{e}(t) =
    \begin{bmatrix}
    x(t) - x_0 \\
    y(t) - y_0
    \end{bmatrix},
\end{equation}
where $(x_0,y_0)$ is the reference centroid corresponding to the VVC center. This error signal is then passed to a proportional-integral-derivative (PID) controller. The PID control law is
\begin{equation}
    u(t) = K_p e(t) + K_i \int_0^t e(\tau)\,d\tau + K_d \frac{de(t)}{dt},
\end{equation}
where $u(t)$ is the controller output, $e(t)$ is the centroid error, and $K_p$, $K_i$, and $K_d$ are the proportional, integral, and derivative gains. In our implementation, the controller is applied independently to the measured $x$ and $y$ centroid errors. The proportional gain is set to 0.15 and the integral gain to 0.002. The derivative gain was set to zero, so the final controller is a PI controller. To convert centroid corrections into TTM commands, we calibrate the response of the LOWFS centroid position to small TTM motions. Because there can be cross-coupling between the X and Y axis on the TTM, we model the response with a $2\times2$ interaction matrix,
\begin{equation}
    \begin{bmatrix}
    \Delta x \\
    \Delta y
    \end{bmatrix}
    =
    J
    \begin{bmatrix}
    \Delta B \\
    \Delta A
    \end{bmatrix},
    \qquad
    J =
    \begin{bmatrix}
    \partial x / \partial B & \partial x / \partial A \\
    \partial y / \partial B & \partial y / \partial A
    \end{bmatrix}.
\end{equation}
$\Delta A$ and $\Delta B$ are small TTM actuator offsets, and $\Delta x$ and $\Delta y$ are the resulting centroid shifts measured on the FLIR. We determine $J$ by applying small calibration pokes of 0.05 mrad to each TTM axis and measuring the corresponding centroid displacement. The inverse matrix $J^{-1}$ is then used during closed loop operation to convert the desired LOWFS centroid correction into TTM actuator commands.

\subsubsection{Dark Hole Digging with EFC}

Alignment alone is not sufficient to reach levels of $10^{-8}$ on the testbed. In order to dig a dark hole where higher contrast can be achieved, we implement an EFC \cite{Giveon2007} routine using CATKit2 on HCST. On HCST, the dark hole is a D-shaped region spanning 4.5--10 $\lambda/D$ (the previous FALCO baseline was 3.5--10$\lambda/D$). EFC is a model-based approach to sense and minimize the electric field in a region on the SCICAM. At each iteration, the focal plane electric field is estimated using pairwise probing. A set of known DM probe shapes is applied with positive and negative signs, and the resulting intensity differences are used to infer the complex field in the dark hole. Given this estimate, the EFC update is computed from the linearized model
\[
\Delta E \approx G \Delta u,
\]
where \(G\) is the Jacobian mapping DM actuator commands to focal-plane electric-field changes and \(\Delta u\) is the incremental DM command. We solve for \(\Delta u\) using a regularized least squares controller and apply the correction with a scalar gain to maintain stable closed-loop behavior on the testbed.

The measured contrast is computed from dark-subtracted coronagraphic images normalized by the direct PSF peak and corrected for exposure time,
\[
C = \frac{(I_\mathrm{coron}-I_\mathrm{dark})t_\mathrm{direct}}
{I_{\mathrm{direct,peak}}t_\mathrm{coron}}.
\]
The contrast reported for each iteration is the mean of \(C\) over the D-shaped dark hole region.
\section{Experimental Results}
Here we present our results on the implementation of the phase retrieval, LOWFS tip-tilt control and EFC on HCST using the CATKit2 software architecture. 


\subsection{Phase Retrieval}

\begin{figure}[ht!]
\begin{center}
\begin{tabular}{c}
\includegraphics[width=0.9\textwidth]{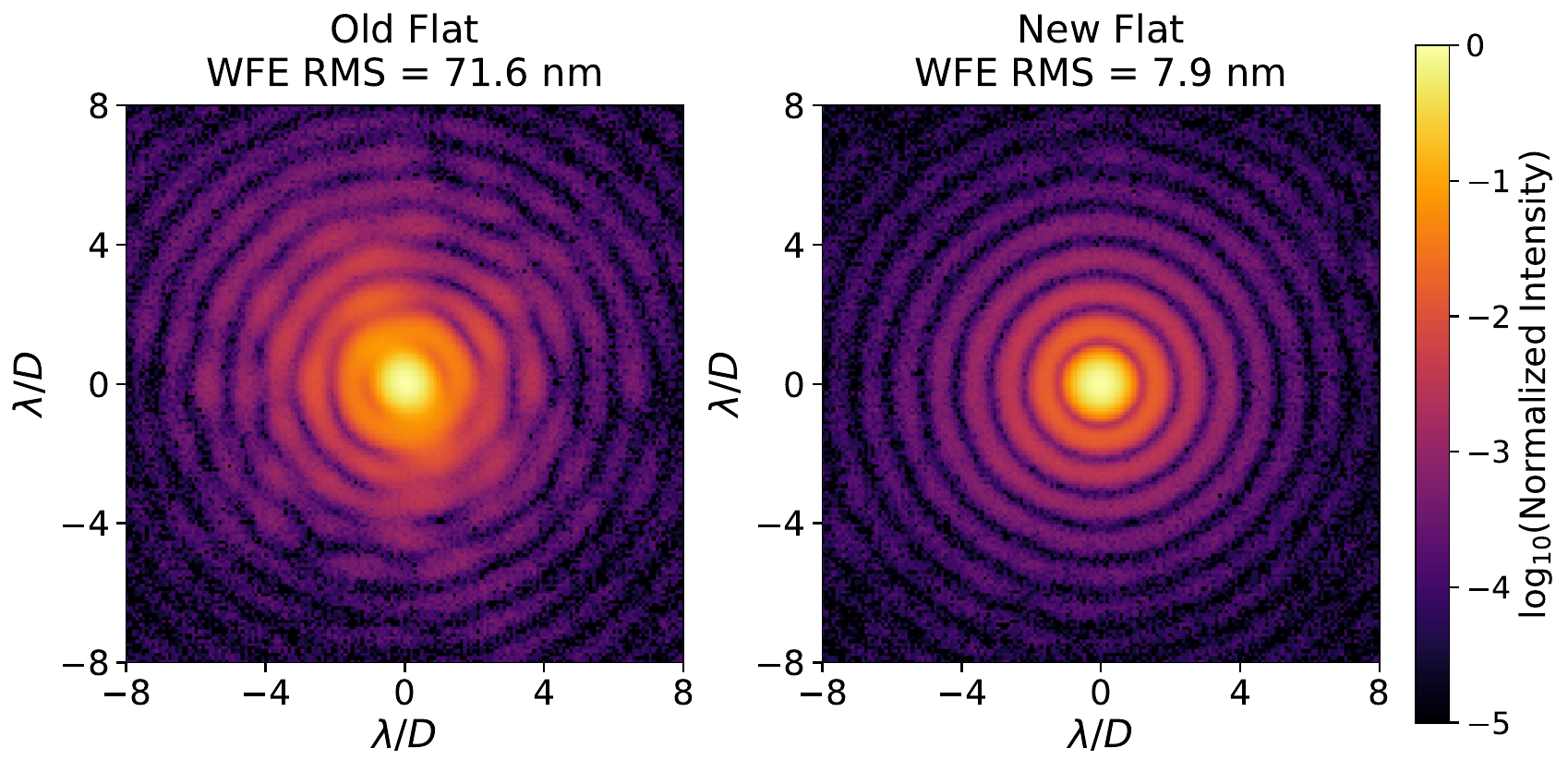}
\end{tabular}
\end{center}
\caption[LOWFS and SCICAM PSFs]
{\label{fig:psfs-pr} Left: The old flat map prior to our phase retrieval algorithm, with a measured WFE RMS of 71.6 nm. Right: Our new flat map after 6 rounds of GS with a measured WFE RMS of 7.9 nm. Sending this new flat map is now the starting point for any routine operations on HCST, including the tip-tilt loop and EFC runs.}
\end{figure}

We quantify the residual aberration using the root-mean-square wavefront error (WFE RMS), defined as the standard deviation of the optical path difference over the illuminated pupil. Lower WFE RMS corresponds to a wavefront closer to the ideal reference surface. Our phase retrieval algorithm using 6 rounds of GS with a defocus offset to break sign ambiguity takes our PSF from  WFE RMS from 71.6 to 7.9 nm (see Figure \ref{fig:psfs-pr}), which corresponds to a $>$9$\times$ reduction in the starting WFE for the SCICAM. The full process from testbed initialization to final PSF takes 18 minutes. The DM surface displacement was estimated using a local voltage-to-height calibration of 1.13 $\mu\mathrm{m}/\mathrm{V}$, a value obtained from previous calibration on HCST. Because the MEMS actuator response is nonlinear over its full command range, this factor does not represent a global stroke calibration.

\begin{table}[ht!]
\begin{center}
\begin{tabular}{c c}
\hline
GS Round & WFE RMS (nm) \\
\hline
1 & 71.6 \\
2 & 22.1 \\
3 & 13.6 \\
4 & 10.4 \\
5 & 9.4 \\
6 & 7.9 \\
\hline
DM surface PV after correction & 225.9 \\
\hline
\end{tabular}
\end{center}
\caption[GS convergence]
{\label{tab:gs_convergence} Wavefront error (WFE) RMS measured during each Gerchberg--Saxton (GS) correction round.  The final DM surface peak-to-valley (PV) is 225.9 nm.}
\end{table}
\subsection{Tip-Tilt Tracking and Control off of the VVC: Performance Results}
In this section, we present our tip-tilt tracking and control results using the LOWFS system at the reflection of the VVC. 
\subsubsection{Environmental Data} \label{environmental}
\begin{figure} [ht!]
   \begin{center}
   \begin{tabular}{c}
   \includegraphics[width = \textwidth]{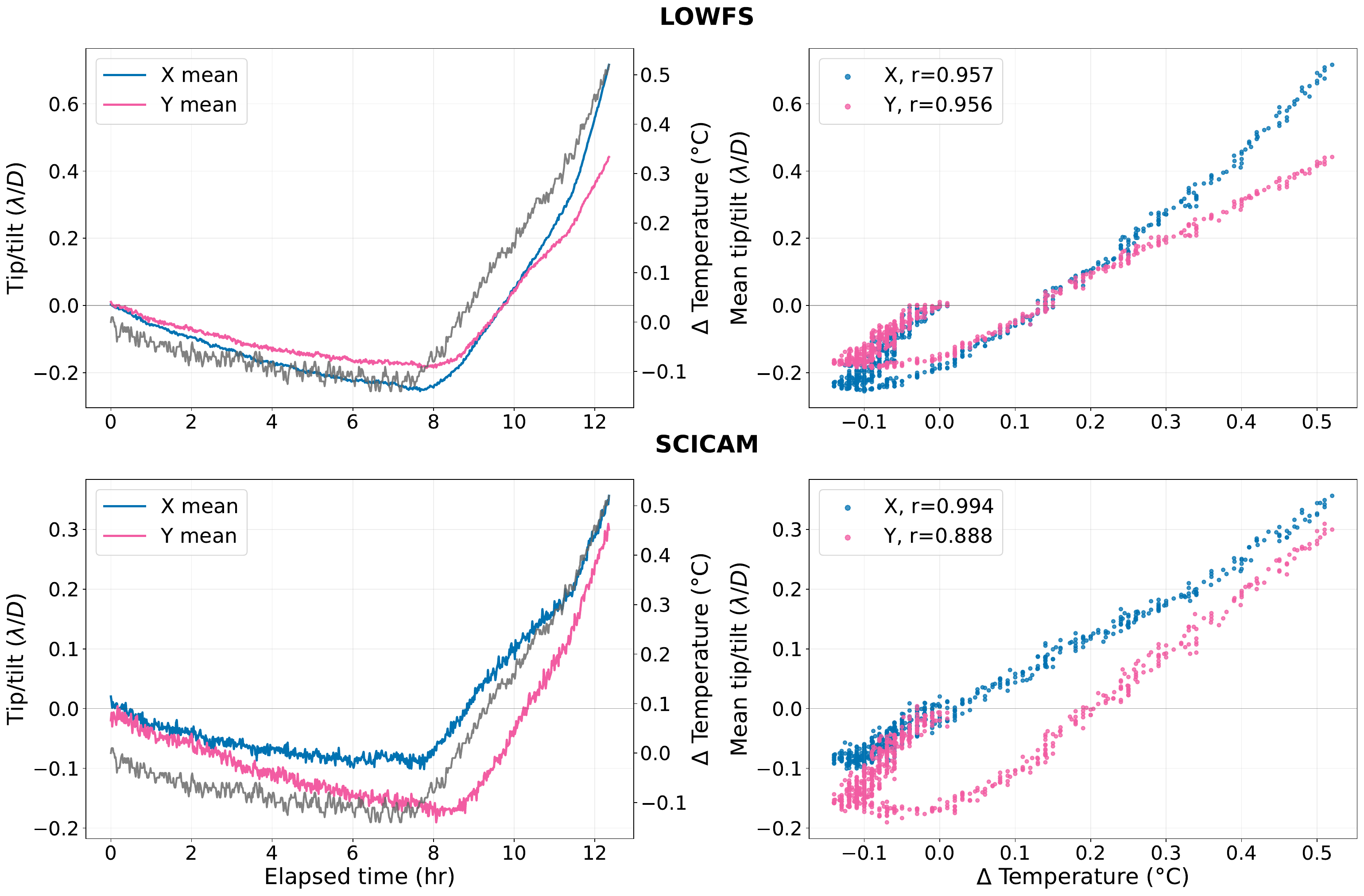}
   \end{tabular}
   \end{center}
   \caption[example] 
   {\label{fig:tt-temp} Open loop tracking of the LOWFS and SCICAM centroids show a correlation between temperature variation and drift on HCST. Left column: tip-tilt offset from reference (0.0) over a 12 hour run for both x-axis (blue) and y-axis (pink), plotted with temperature variations during the same time period for comparison (gray). At around 8 hours after tracking, a temperature increase on the testbed leads to an increase in tip-tilt drift. Right column: Correlation between temperature variation and mean tip-tilt drift. The correlation coefficient, computed here as the Pearson r, shows a correlation factor of near 1, confirming the drift is likely caused by temperature changes on the testbed. }
   \end{figure}

We monitor the temperature and humidity on HCST using small sensors inside the bench to determine whether variations in these environmental parameters are correlated with drift of the PSF. To test this, we tracked the PSF centroid for 12 hours in open loop while simultaneously recording the temperature and humidity inside the testbed. The environmental data is reported at 0.1 Hz, and the tip-tilt data is reported at 100 Hz. For that reason, we average the data from both components every 1 minute in order to assess potential correlations. Figure \ref{fig:tt-temp} shows the measured tip-tilt drift and temperature variation, along with their correlations.

To quantify the correlation between the environmental variations and the measured tip-tilt drift, we compute the Pearson correlation coefficient, defined as
\begin{equation}
r_{xy} =
\frac{\sum_{i=1}^{N} (x_i - \bar{x})(y_i - \bar{y})}
{\sqrt{\sum_{i=1}^{N} (x_i - \bar{x})^2}
 \sqrt{\sum_{i=1}^{N} (y_i - \bar{y})^2}},
\end{equation}
where $x_i$ is the environmental variation, either $\Delta T$ or $\Delta H$, and $y_i$ is the measured one-minute mean tip-tilt centroid position. The quantities $\bar{x}$ and $\bar{y}$ are the corresponding sample means. A value of $r=1$ indicates a perfect positive linear correlation. \par
We find significant correlations ($>$0.888) between the drift and temperature during the 12 hour run. The temperature varied $>$0.5$^{\circ}$ overnight, with a sudden temperature increase around 8 hours into tracking leading to a reversal in the tip-tilt drift direction. For humidity, there is also a level of correlation ($>0.69$ for every variable), but not as high as for temperature, where levels of correlation consistently reached $\gtrsim 0.9$. The humidity varied by 0.4$\%$ during our 12 hour run. From these results, we conclude that drift on HCST is mostly due to thermal and humidity changes on the testbed.

\subsubsection{Tip-Tilt Control}
\begin{figure}
  \begin{center}
    \centering
    \centering{{\includegraphics[width=0.8\textwidth]{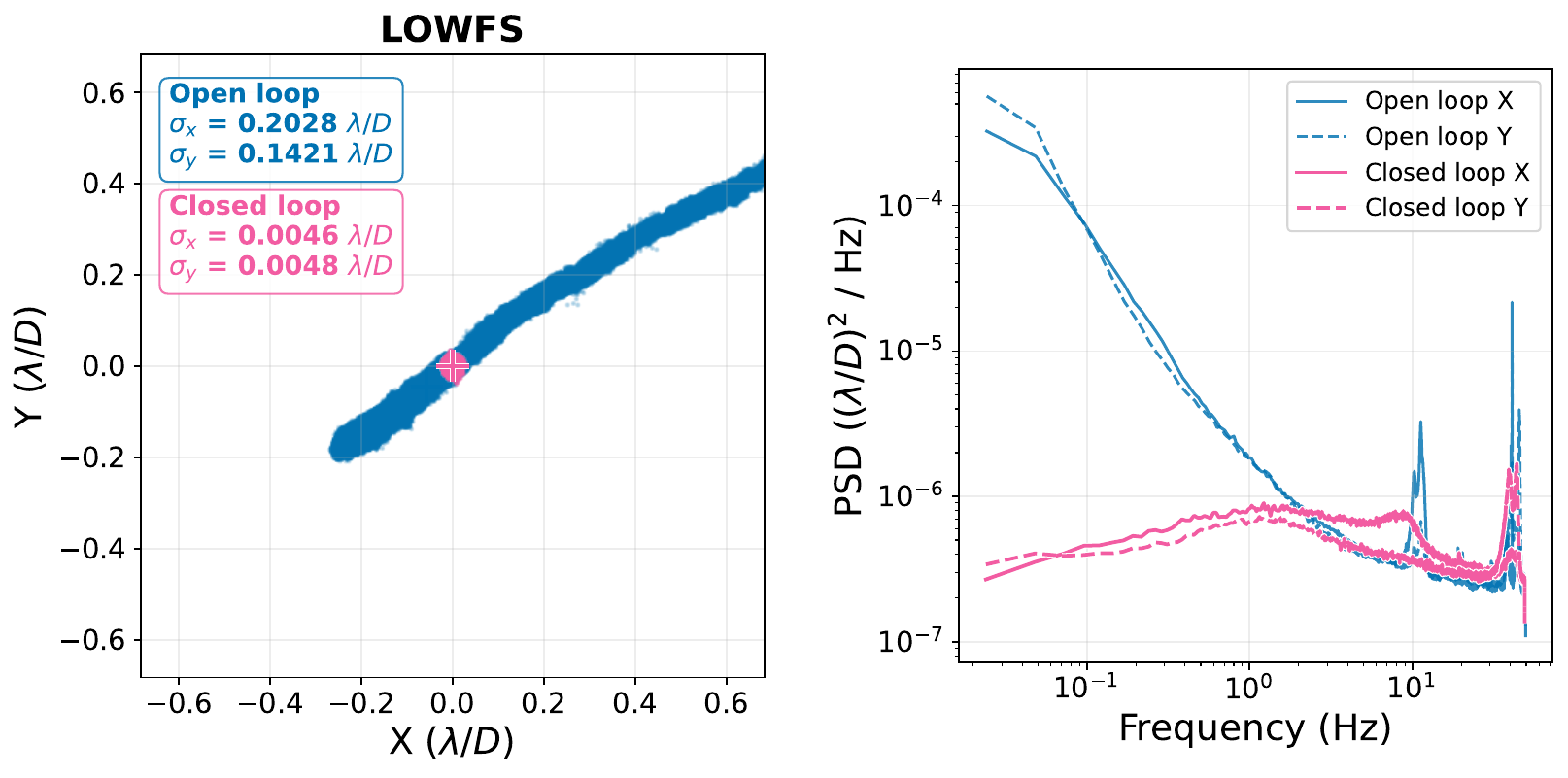} }}%
    \qquad
    \centering{{\includegraphics[width=0.8\textwidth]{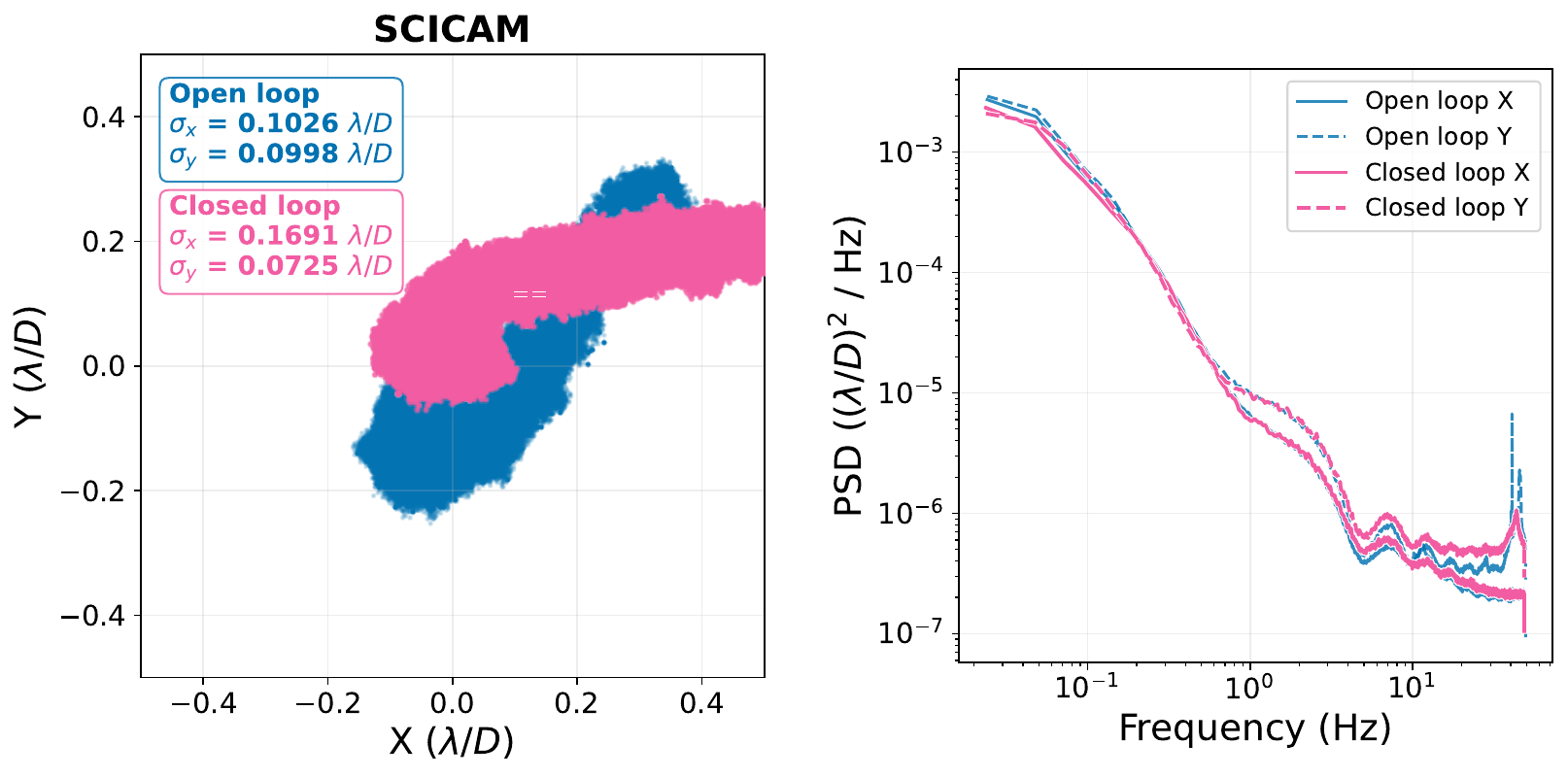} }}%
\caption{Open and closed loop tip/tilt stability measured on the LOWFS and SCICAM. 
The left panels show the centroid error distributions in $\lambda/D$, while the right panels show the corresponding power spectral densities (PSDs) for the $x$ and $y$ axes. 
Closing the loop on the LOWFS suppresses low frequency drift below $1~\mathrm{Hz}$ by more than two orders of magnitude and maintains a pointing precision of $<0.005~\lambda/D$. 
The SCICAM does not show the same reduction, suggesting that the remaining downstream drift is dominated by non common-path motion between the LOWFS the SCICAM. Drift at the LOWFS is upstream of the VVC, generally the main cause of degraded contrast}.
\label{fig:ttloop}
  \end{center}
\end{figure}

   After finding significant drift of our PSF due to environmental parameters, we implement the PID control loop presented in Section \ref{pid} on the LOWFS, while simultaneously tracking the centroid on the SCICAM. We run the loop for 12 hours and compare to our open loop results from Section \ref{environmental}. Our results are shown in Figure~\ref{fig:ttloop}. We find a significant improvement of more than two orders of magnitude in the LOWFS drift at frequencies below $1~\mathrm{Hz}$ along both axes, with a consistent pointing precision of $<0.005~\lambda/D$. 
   \par
   The closed loop correction does not appear to improve the drift measured on the SCICAM, whose PSD remains essentially unchanged between the open and closed loop cases, and drift levels remain comparable to the open loop. Since there are many optics between the LOWFS and the SCICAM, there is a possibility of non-common path aberrations between the two PSFs. However, the purpose of the LOWFS control loop is to eliminate drift upstream of the VVC, since pointing errors upstream of the coronagraph are among the dominant contributors to degraded contrast and contrast stability. For this reason, the effectiveness of the control loop can ultimately only be assessed when it is run concurrently with EFC.

\subsubsection{Dark Hole Digging + LOWFS on reflection of the VVC}

Here we present the results of our first implementation of EFC on HCST with CATKit2, and its parallel operation with the LOWFS. Our goal is to assess how much the tip-tilt drift found in the previous section can potentially affect our contrast stability.
\par
Our EFC implementation in CATKit2 reaches contrast levels of $\sim$5e-8 in a dark hole region between 4.5 -- 10 $\lambda/D$ at 780 $\pm$ 10 nm wavelengths. We use three pairs of sinc function probes for redudancy in order to estimate the local electric field in the SCICAM. HCST's previous baseline with FALCO for a dark hole region of 3.5 -- 10 $\lambda/D$ was 1e-8 \cite{Arielle2024}, suggesting our CATKit2 HCST implementation has a mismatch between the optical model of the testbed and the true bench response, which is being investigated. However, we noticed a significant increase in the EFC iteration speed since our switch to CATKit2: previously, 50 EFC iterations with FALCO would take $\sim$ 30 minutes, while with CATKit2 it takes $<$15 minutes, a factor of 2 increase. Due to its service-based architecture, CATKit2 should be able to dig a dark hole within a few seconds, rather than minutes \cite{por_2026}, suggesting there are remaining overheads in our current implementation. Potential overheads remaining involve a complex wrapper around the EFC loop, in particular for sending probes. In the future, our goal is to turn the entire EFC loop (and LOWFS loop) into a standalone CATKit2 service, essentially eliminating any additional overheads. 
\par
   \begin{figure} [ht!]
   \begin{center}
   \begin{tabular}{c}
   \includegraphics[width = \textwidth]{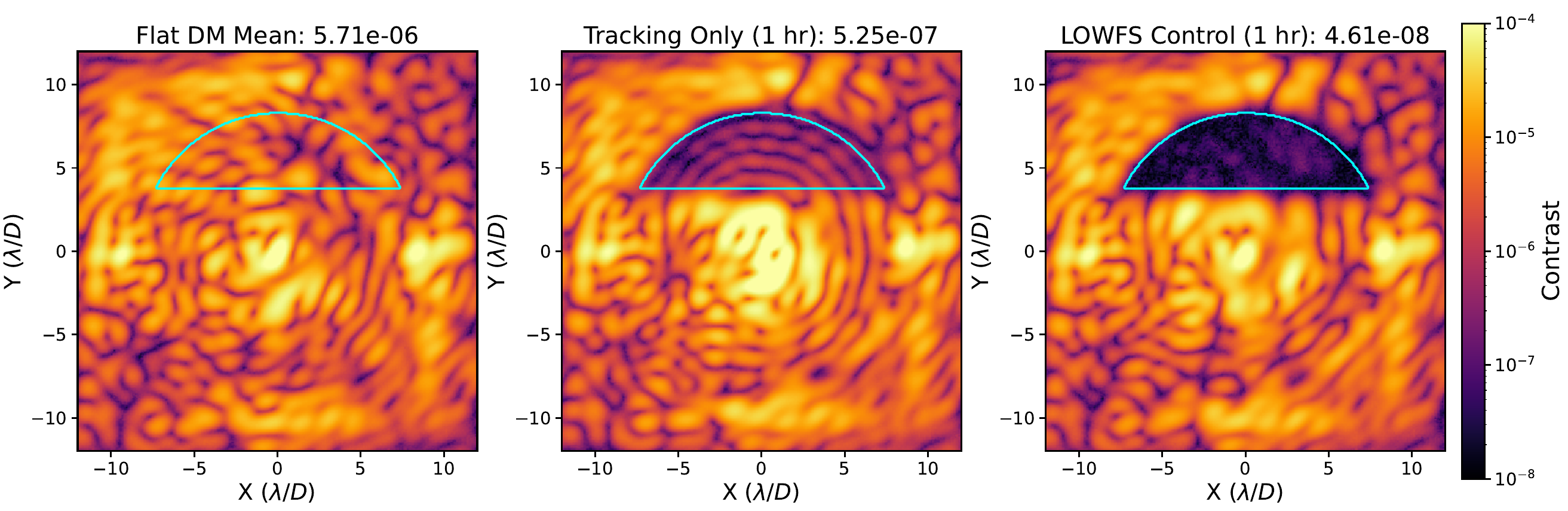}
   \end{tabular}
   \end{center}
   \caption[example] 
   {\label{fig:DH-Image} Left: The dark hole region is highlighted, where the speckle pattern produces a contrast of $\sim 5.7\times10^{-6}$ between 4.5--10$\lambda$/D. Middle: Uncontrolled LOWFS tracking, where the centroid offset is measured relative to the start of the run but not corrected, shows a $\sim 10\times$ contrast degradation over one hour, from $5.5\times10^{-8}$ to $5.25\times10^{-7}$ (blue curve in Figure \ref{fig:contrast-stability}). Right: With LOWFS control using light reflected from the VVC, the dark hole contrast is maintained at $\sim 4.5$--$5\times10^{-8}$ by the end of the hour (pink curve in Figure \ref{fig:contrast-stability}).}
   \end{figure} 

CATKit2 allows for multiple complex loops to occur simultaneously on the testbed. We implement a combination of EFC + tip-tilt control in order to monitor/control the PSF pointing on the FPM with the LOWFS and TTM, while simultaneously using the SCICAM and the DM to dig a dark hole with EFC. In order to assess stability of our tip-tilt control while EFC is operational, we run 30 iterations of EFC and continuously monitor the contrast in the dark hole region for 1 hour. We run this test for two cases: LOWFS tracking (open loop) and control (closed loop). Similarly to the previous section, we find a large drift ($>$0.1 $\lambda/D$) in the open loop case, while the closed loop remains $\sim0.005 \lambda/D$ at all times. 
\par
\begin{figure} [ht!]
   \begin{center}
   \begin{tabular}{c}
   \includegraphics[width = \textwidth]{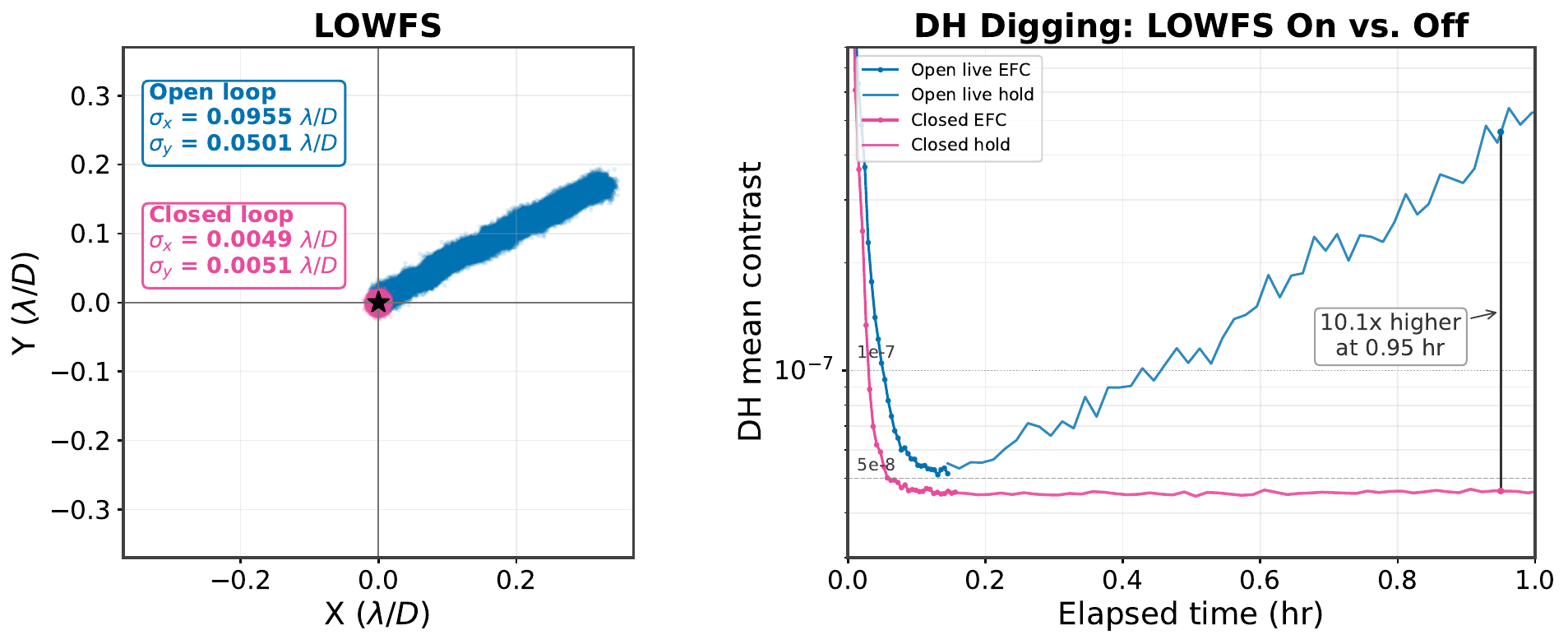}
   \end{tabular}
   \end{center}
   \caption[example] 
   {\label{fig:contrast-stability}Left: LOWFS pointing stability for closed loop (pink) and open loop (blue). The open loop drifts by $>0.2\lambda/D$, which we found can be correlated with temperature variations in the lab. Right: mean contrast in the dark hole (DH) region of 4.5--10$\lambda/D$ over one hour, after 30 iterations of EFC, with and without the LOWFS control at the reflection of the VVC. With LOWFS control (pink), the contrast levels are maintained at $<5\times10^{-8}$, while without it (blue) the contrast degrades quickly from the end of the 30th iteration of EFC, reaching a contrast degradation of $>10\times$ within 1 hour.}
   \end{figure}

We find that the drift in the open loop case has significant impact on our dark hole stability. Despite similar levels of contrast after 30 iterations (4.5 vs. 5.5e-8) in the open loop case, a significant drift of $\sim$ 0.2 $\lambda/D$ leads to an over one order of magnitude contrast loss in the dark hole region (reported in Figures \ref{fig:DH-Image} and \ref{fig:contrast-stability}) in just one hour of tracking. Implementing the LOWFS system thus demonstrates that pointing stabilization can occur at the reflection of the VVC, and shows the important role it can play in contrast stability. However, the presence of the drift on the testbed is not ideal for long-term operations, even with LOWFS control, and it was not always present in the lab \cite{Arielle2024}. We believe the tip-tilt mirror itself is drifting due to temperature changes, since it is the only component that changed from 2024 to now. We are currently investigating whether the drift is due to a software issue (e.g., servos not being on, despite appearing as ``on'' the CATKit2 software) or a hardware issue (e.g., loose mounting).
\par

\section{Discussion \& Future Work}

We have presented initial experimental results of a LOWFS tip-tilt control system operating
at the reflection of a charge 8 VVC on HCST, implemented within the new CATKit2
software architecture. Our phase retrieval routine reduces the SCICAM wavefront
error from 71.6 to 7.9 nm RMS, a $>$9$\times$ improvement, and
produces the flat map that now seeds all routine operations on the testbed.
Closing the tip-tilt loop on light reflected from the VVC suppresses low frequency drift by more than two orders of magnitude, taking residual pointing from
$\sigma \approx 0.20/0.14~\lambda/D$ in open loop to $\lesssim 0.005~\lambda/D$ in closed loop. Most importantly, when EFC and the LOWFS loop are run concurrently,
the closed loop holds pointing near $0.005~\lambda/D$ and maintains a dark hole
contrast of $\sim$5$\times10^{-8}$, whereas in open loop a
drift of $\sim$0.2--0.4~$\lambda/D$ degrades the contrast by more than an order of
magnitude over the same interval. Together these results demonstrate for the
first time that pointing can be sensed and stabilized at the reflection of a VVC,
and that doing so plays a meaningful role in coronagraphic contrast stability.

Several tests remain to be conducted. First, we will continue to trace the
sources of drift on the bench beyond temperature, in particular isolating whether
the tip-tilt mirror itself is drifting due to thermal effects or a software or hardware issue, since it is the dominant change to the testbed since 2024 when drift levels were minimal \cite{Arielle2024}.
Second, we will quantify and reduce the mismatch between our optical model and the
true bench response that currently limits our CATKit2 EFC contrast relative to the
prior FALCO baseline, focusing on DM registration and orientation and Lyot stop alignment. The goal is to establish a baseline previously found on HCST prior to the tip-tilt mirror installation and transition to CATKit2, which was a mean dark hole contrast of 1e-8 between 3.5–10$\lambda/D$.
 Third, we plan to reimplement the EFC loop (and the LOWFS loop) as standalone CATKit2 services to eliminate any remaining software overhead and enable dark hole digging at much higher cadence alongside
LOWFS control, approaching the few-second dark-hole-digging times that CATKit2 should allow \cite{por_2024_11395554}. Once these are resolved, we will test 
apodizers\cite{SusanJATIS} for HWO and update the DM electronics, both of which we expect to further validate the LOWFS+VVC architecture as a candidate for HWO.
\acknowledgments 
C.D.O thankfully acknowledges the support of the B. Thomas Soifer Fellowship at Caltech.

\bibliography{report} 
\bibliographystyle{spiebib} 

\end{document}